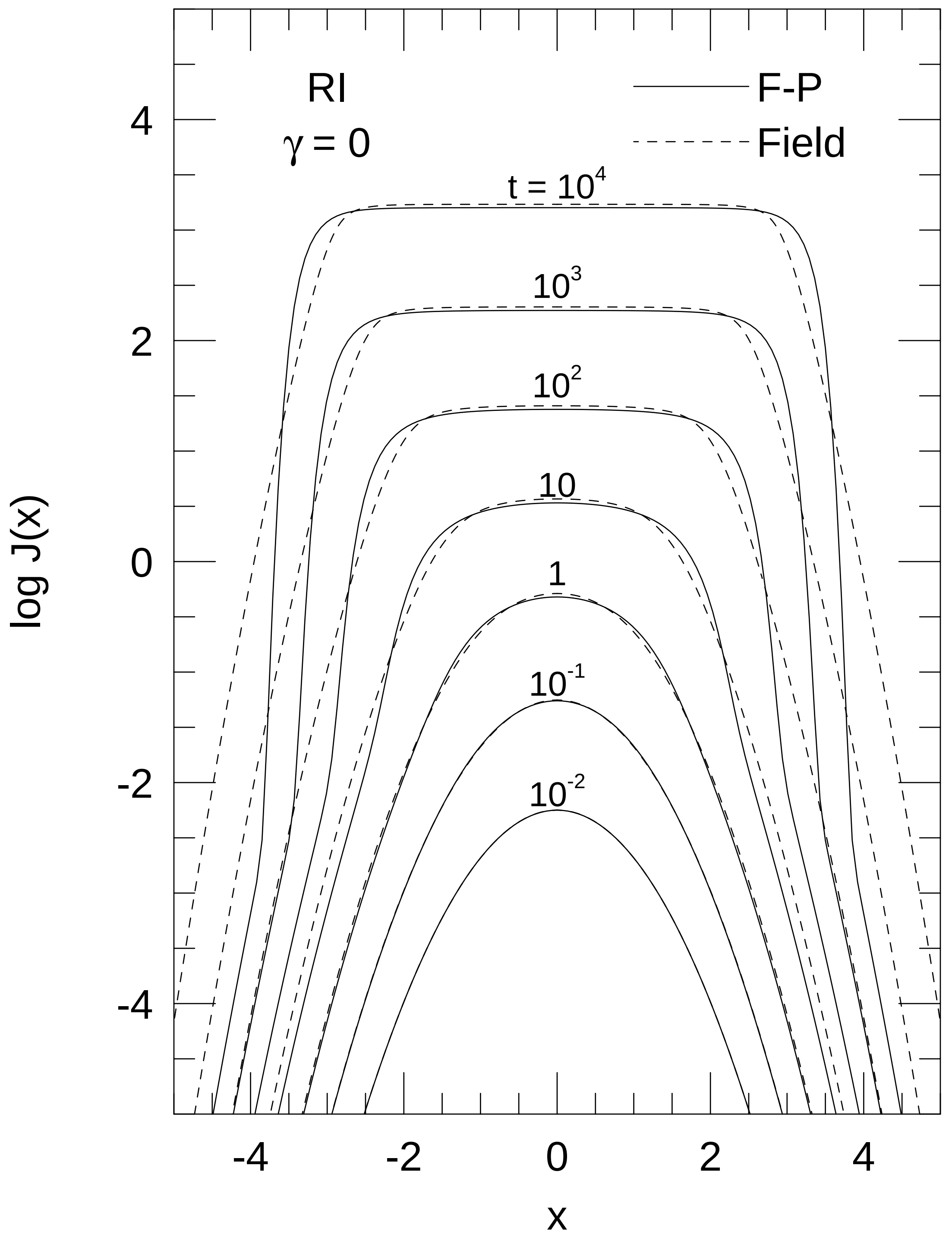

# THE TIME DEVELOPMENT OF A RESONANCE LINE

# IN THE EXPANDING UNIVERSE

George B. Rybicki

Ian P. Dell'Antonio

Harvard-Smithsonian Center for Astrophysics

Cambridge, MA 02138

ABSTRACT

The time-dependent spectral profile of a resonance line in a homogeneous expanding medium is studied by numerically solving an improved Fokker-Planck diffusion equation. The solutions are used to determine the time required to reach a quasi-static solution near the line center. A simple scaling law for this relaxation time is derived and is fitted to the numerical results. The results are applied to the case of Lyman alpha scattering during primordial recombination of hydrogen. For a wide range of cosmological models it is found that the relaxation times are smaller than the recombination timescale, although not by a very large factor. Thus the standard assumption of a quasi-static solution in cosmological recombination calculations is reasonably valid, and should not cause substantial errors in the solutions.

## 1. INTRODUCTION

The time development of resonance line profiles has evoked interest over the years because it plays a role in a variety of astrophysical problems, from the study of the resonance excitation of the hydrogen 21-cm line (Field, 1959) to analyses of the emission of x-ray resonance lines (Basko 1978). Previous studies have examined the case of an infinite, homogeneous medium that is static, that is, with no macroscopic velocity fields. These studies have considered the scattering to be described by a variety of redistribution mechanisms (using the notation of Hummer 1962), including $R_I$ redistribution by Field (1959), complete redistribution over pure doppler profiles by Ivanov (1967), complete redistribution over pure Lorentz profiles by Basko (1978), and type $R_{II}$ redistribution in the Voigt wings by Basko (1978). These studies of static media have shown that the detailed time development is very strongly dependent on the type of redistribution mechanism operating in the particular resonance line.

A non-static case of considerable interest is that of a infinite homogeneous expanding medium, since this is a good approximation to the conditions of the universe at the time of hydrogen recombination. Provided that the expansion is sufficiently slow, one expects that the time-dependent line radiation field will eventually approach an equilibrium (quasi-static) configuration. These equilibrium solutions have been extensively studied (see, e.g., Rybicki and Hummer 1992 and references therein). In previous studies of the cosmological hydrogen recombination epoch (Peebles 1968; Zel'dovich, Kurt and Sunyaev 1969; Krolik 1989, 1990), it has been assumed that the Lyman alpha line profile rapidly reaches this quasi-static configuration, and only the quasi-static solution has then been considered in solving the recombination problem.

The question of whether the Lyman alpha line is in quasi-static equilibrium is very important for some aspects of cosmological recombination. The reasons are similar to those for the analogous problem of recombination in diffuse nebulae (see, e.g., Osterbrock 1989), where the trapping of Lyman alpha radiation is an important process in controlling the recombination rates, and it is the lack or presence of such trapping that distinguishes between Case A and Case B recombination. In the cosmological case, if there is insufficient time to build up the quasi-static solution, then the trapping is less effective than otherwise, and recombination will be speeded up; this would lead, e.g., to a lower residual ionization of the IGM, which would affect subsequent molecule formation (Palla, Salpeter, & Stahler 1983; Lepp & Shull 1984).

The present investigation was motivated by some preliminary, rough estimates of the relaxation time for the Lyman alpha line profile, which indicated that it may be comparable with the cosmological expansion timescale. If the two timescales were equal, or indeed if the relaxation timescale were longer than the expansion timescale, then the previous recombination calculations would be in error. The purpose of this paper is to examine in detail the relation between these two timescales.

In §2 the basic equations describing the time dependence of a resonance line profile in an expanding universe will be derived. The redistribution mechanism is assumed to be $R_{II}$, appropriate to the hydrogen Lyman alpha line during the cosmological recombination epoch. The Fokker-Planck method is used here for treating $R_{II}$ redistribution. The standard formulations of this method are due to Harrington (1973) and Basko (1978), following the pioneering work of Unno (1955). In this paper we derive and use a new version of the Fokker-Planck approximation that has improved behavior near the Doppler core (see the Appendix). In §3 analytic results for time dependent development of line radiation fields in static media are reviewed, and some new results for the uniform expansion case is given. In §4 the numerical method is described and results are given for the relaxation time. A simple scaling law is derived that fits the numerical results for the relaxation times quite well. These results are then applied to the cosmological recombination problem. Section 5 contains an evaluation of the importance of some physical effects not treated in our equations. Section 6 gives our conclusions.

## 2. BASIC EQUATIONS

For resonance lines with sharp lower levels, such as the Lyman transitions in hydrogen at low densities, collisional broadening is negligible and natural broadening dominates. In this case the absorption line profile is a Voigt function, but the emission profile depends on the spectrum of the absorbed radiation. In the rest frame of the atom, the re-emitted (scattered) radiation is coherent, but in the observer's frame the scattering is noncoherent, because of the Doppler effect of the atom's thermal motion. This corresponds to type $R_{II}$ redistribution (Hummer 1962).

The transfer equation for resonance scattering in an infinite, static, homogeneous and isotropic medium can be written as

$$\frac{1}{c}\frac{\partial J_p(\nu,t_p)}{\partial t_p} = -\chi\varphi(\nu)J_p(\nu,t_p) + \chi\int \mathcal{R}_{II}(\nu,\nu')J_p(\nu',t_p)\,d\nu' + C_p(t_p)\varphi(\nu), \qquad (1)$$

where $J_p(\nu,t_p)$ is the mean intensity (defined here in terms of photon numbers rather than energies) as a function of frequency $\nu$ and time $t_p$. [The subscript "p" is used here to denote "physical" variables that will be replaced by rescaled versions later.] $\mathcal{R}_{II}(\nu,\nu')$ is the angle-averaged redistribution function describing how the photon absorbed at $\nu'$ is re-emitted at $\nu$. (The use of the angle-averaged form is permissible because of the isotropy of the radiation field.) Neglecting stimulated emission, the line absorption coefficient is given by $\chi\varphi(\nu)$, where

$$\chi = \frac{h\nu_0}{4\pi} n_1 B_{12} = \frac{\pi e^2}{m_e c} n_1 f_{12}, \tag{2}$$

and $\varphi(\nu)$ is the line profile function (with normalization $\int \varphi(\nu)\, d\nu = 1$), which is assumed to be a Voigt profile. In Eq. (2), $e$ and $m_e$ are the electron charge and mass; $c$ is the speed of light; $n_1$ is the number density of the lower state; and $B_{12}$ and $f_{12}$ are the associated Einstein coefficient and oscillator strength. The function $C_p(t_p)$ describes the rate of creation of new resonance photons due to the radiative cascade following recombination; the actual form of $C_p(t_p)$ can only be determined on the basis of detailed physical models of the recombination process. We have assumed that the newly created photons are emitted with the profile $\varphi$, which is reasonable for the radiative cascade model.

Equation (1) applies to a static medium. In an isotropically expanding infinite medium, this equation needs to be modified by adding terms to the left hand side corresponding to the density change and redshifting of the radiation. Following (Peebles, 1968), these terms can be written as

$$2\frac{\dot{R}}{R} J_p(\nu, t_p) - \nu \frac{\dot{R}}{R} \frac{\partial J_p(\nu, t_p)}{\partial \nu}, \tag{3}$$

where the $R(t_p)$ is the cosmological scale factor; the value of $R(t_p)$ at some standard epoch (in our case, the present) will be denoted by $R_0$. [Note that the factor 2 (not 3) appears in Eq. (3), because of our definition of $J_p$ in terms of photon numbers rather than energies.] Equation (1) then becomes

$$\frac{1}{c}\frac{\partial J_p(\nu, t_p)}{\partial t_p} + 2\frac{\dot{R}}{R} J_p(\nu, t_p) - \nu \frac{\dot{R}}{R} \frac{\partial J_p(\nu, t_p)}{\partial \nu}$$
$$= -\chi\varphi(\nu) J_p(\nu, t_p) + \chi \int \mathcal{R}_{II}(\nu, \nu') J_p(\nu', t_p)\, d\nu' + C_p(t_p)\varphi(\nu). \tag{4}$$

It is convenient to change to a dimensionless frequency variable $x$ which measures the frequency relative to line center $\nu_0$ in units of the Doppler width, i.e. $x = (\nu - \nu_0)/\Delta\nu_D$ where $\Delta\nu_D = \nu_o v_T/c$. Here $c$ is the speed of light and $v_T$ is the thermal speed of the atom, given by $v_T = (2k_B T/M)^{1/2}$, where $k_B$ is Boltzmann's constant, $T$ is the temperature, and $M$ is the mass of the atom. For cases of interest here the thermal speeds of the atoms are very non-relativistic, $v_T/c \ll 1$, so that Eq. (4) becomes

$$\frac{\partial J_p(x, t_p)}{\partial t_p} + 2H J_p(x, t_p) - \frac{cH}{v_T}\frac{\partial J_p}{\partial x}$$
$$= -kc\phi(x) J_p(x, t_p) + kc \int R_{II}(x, x') J_p(x', t_p) dx' + C_p(t_p)\phi(x), \tag{5}$$

where $H(t_p)$ is the local Hubble constant $H(t_p) = \dot{R}(t_p)/R(t_p)$. The line absorption coefficient is now given by $k\phi(x)$, where $k = \chi/\Delta\nu_D$ and $\phi(x)$ is the standard, normalized Voigt function, and $R_{II}(x, x')$ is the standard normalized redistribution function. (see, e.g., Rybicki and Hummer 1992, Eqs. [3.1]–[3.5]). Actually we should write $\phi(x, a)$ instead of $\phi(x)$, since the Voigt profile depends on the Voigt parameter $a$, which is the ratio of Doppler to natural width of the line $a = A_{21}/4\pi\Delta\nu_D$, where $A_{21}$ is the Einstein $A$-coefficient for the transition. However, for notational simplicity we suppress the Voigt parameter in $\phi(x)$. For cases of interest here, the Voigt profile is dominated by the Doppler core near line center, where $\phi(x) \approx \pi^{-1/2}\exp(-x^2)$, so that $k$ is approximately $\pi^{1/2}$ times the line center absorption coefficient. It is convenient to define a new dimensionless time variable

$$t = kct_p, \tag{6}$$

which measures time in units of the mean free time, defined in terms of the integrated line absorption coefficient $k$. We also introduce the rescaled variables

$$J(x,t) = (R(t_p)/R_0)^2 J_p(x,t_p), \qquad C(t) = (R(t_p)/R_0)^2 C_p(t_p). \tag{7}$$

The transfer equation then takes the form

$$\frac{\partial J}{\partial t} = -\phi(x)J(x,t) + \int R_{II}(x,x')J(x',t)dx' + \gamma\frac{\partial J}{\partial x} + C(t)\phi(x). \tag{8}$$

Here we have introduced the well-known *Sobolev parameter*

$$\gamma = \frac{H}{v_T k} = \frac{8\pi H}{3A_{21}\lambda^3 n_1}, \tag{9}$$

which characterizes the expansion rate. The second form, in terms of the Einstein $A$-coefficient and wavelength $\lambda$ for the Lyman alpha transition, follows from the Einstein relations and the ratio $g_2/g_1 = 3$ for the statistical weights. Physically, $\gamma$ is the ratio of the mean free path (defined in terms of $k$) to the distance over which the expansion of the medium induces a velocity difference equal to the thermal velocity. Another interpretation is that the total optical depth encountered by a photon that redshifts completely through the line is $\tau_S \equiv \gamma^{-1}$, sometimes called the *Sobolev optical depth*. We see that the non-static equation is now of the same form as the static equation, except for the additional term $\gamma(\partial J/\partial x)$, which describes the redshifting of photons due to the general expansion of the medium.

The treatment of transfer problems involving the redistribution function $R_{II}$ can be quite difficult, since they involve solving integro-differential equations. Fortunately, assuming the $J(x,t)$ varies sufficiently slowly over an interval $\Delta x \approx 1$, the scattering integral involving $R_{II}$ may be approximated by a differential operator (Harrington 1973; Basko 1978, 1981); this is also known as the *Fokker-Planck approximation*. In the appendix we derive an improved version of the approximation, in which the required moments of the redistribution function are found exactly. Our approximation is

$$\int R_{II}(x,x')J(x')\,dx' = \phi(x)J(x) + \frac{1}{2}\frac{\partial}{\partial x}\left(\phi(x)\frac{\partial J(x)}{\partial x}\right), \tag{10}$$

which differs from the previous versions in that the full Voigt profile is used in the differential term, rather than its asymptotic form $\phi(x) \sim a/\pi x^2$, for $|x| \gg 1$. For this reason our approximation is expected to have improved properties near the Doppler core. Using Eq. (10), the transfer equation (8) now simplifies to

$$\frac{\partial J(x,t)}{\partial t} = \frac{1}{2}\frac{\partial}{\partial x}\left(\phi(x)\frac{\partial J(x,t)}{\partial x}\right) + \gamma\frac{\partial J}{\partial x} + C(t)\phi(x). \tag{11}$$

Apart from the function $C(t)$, this equation depends only on two parameters, the Sobolev parameter $\gamma$ and the Voigt parameter $a$.

It should be noted that the Fokker-Planck approximation to the transfer equation, although computationally simple, has some deficiencies. In particular, it is only a good approximation when $J$ is not a rapidly varying function of $x$ (since we are neglecting the derivatives of order higher than two). The regions in $x$ and $t$ where our solution is not reliable will be discussed later. The formulation of Eq. (11) neglects the recoil of the atom during scattering, but, as we show in §6, this should not seriously affect our results.

## 3. ANALYTIC RESULTS

There are few known analytic results concerning time-dependent line transfer with $R_{II}$ redistribution. We review and generalize the results of Field (1959) and Basko (1979) for static media in sections 3.1 and 3.2. A new exact result for time-dependent coherent scattering in an expanding medium is given in §3.3. Finally, in §3.4, we present a simplified transfer equation for time-dependent $R_{II}$ redistribution, which is applicable for small $\gamma$, and which has simple scaling properties.

### 3.1. *Static Medium; The Field Solutions*

Field (1959) found exact solutions for the time development of a radiation field in a static medium for $R_{II}$ redistribution with $a = 0$ (also called $R_I$ redistribution), which is also a good aproximation when $a \ll 1$ for sufficiently early times. The Field solution for an initial field in the form of a Doppler profile $J_i(x,0) = \pi^{-1/2}\exp(-x^2)$, with zero source, $C = 0$, can be written, after integration by parts and other manipulations,

$$J_i(x,t) = \pi^{-1/2} e^{-x^2} \exp(-te^{-x^2}) + t\int_x^\infty e^{-w^2}\exp(-te^{-w^2})\mathrm{erf}(w)\,dw, \tag{12}$$

where erf denotes the well known error function (Field 1959; Eq. [16c]). The Field solution for a unit, constant source, $C(t) = 1$, with an initially zero radiation field, is given by

$$J_s(x,t) = \pi^{-1/2}\left[1 - \exp(-te^{-x^2})\right] + \int_x^\infty e^{w^2}\left[1 - \left(1 + te^{-w^2}\right)\exp(-te^{-w^2})\right]\mathrm{erf}(w)\,dw, \tag{13}$$

(Field 1959; Eq. [19b]).

### 3.2. *Static Medium; The Basko Solutions*

Basko (1978) found solutions for $R_{II}$ redistribution using a Fokker-Planck approximation due to Harrington (1973). The Basko solution for the development from an initial pulse of radiation, $J(x,0) = \delta(x)$, and zero source, $C(t) = 0$, is,

$$J_i(x,t) = \frac{2}{\Gamma(1/4)\bar{t}^{1/4}} e^{-x^4/\bar{t}}, \tag{14}$$

where $\bar{t}$ is a characteristic time $\bar{t} = 8at/\pi$. Although Basko did not give the result $J_s(x,t)$ for a constant source, $C(t) = 1$, this can be easily derived from the result (14), if we use the *delta-function injection approximation*, $\phi(x) \sim \delta(x)$, for injection of photons in the the source term. In this case, $J_s$ can be expressed as a superposition of pulses at different times, which through a series of straightforward manipulations can be shown to be

$$J_s(x,t) = \int_0^t J_i(x,t')\,dt' = \frac{\pi}{4a\Gamma(1/4)}|x|^3\Gamma(-3/4, x^4/\bar{t}). \tag{15}$$

Here $\Gamma(-3/4, x^4/\bar{t})$ denotes an incomplete gamma function (see, e.g., Abramowitz and Stegun 1964).

The Fokker-Planck approximation as used by Basko has some limitations: it is valid only in the wing regions and it assumes that all the new radiation is injected solely at line center, $x = 0$. These limitations will be clear later when comparing with results from our numerical solutions.

Some of the deficiencies of the Basko approximation could in principle be overcome using the analytic results of Grachev (1988), who gave the Green's function for the injection of source radiation at arbitrary values of frequency. However, the solution is very complicated, involving an integral over Bessel functions. This would have to be further integrated over a Voigt profile and integrated over time to obtain the solution for the constant source problem.

### 3.3. *Expanding Medium; Coherent Scattering*

It is possible to give an analytic solution to the non-static equations for the special case of no diffusion, that is when the Fokker-Planck term is absent. This does not represent a real physical situation, but it provides an interesting and useful reference solution for the more realistic cases of $R_{II}$ redistribution. This case is equivalent to that of *coherent scattering*, where we formally substitute $R_{II}(x,x') = \phi(x)\delta(x-x')$. Without the diffusion term, Eq. (11) becomes,

$$\frac{\partial J}{\partial t} = \gamma\frac{\partial J}{\partial x} + C\phi. \tag{16}$$

This equation can easily be solved by changing from variable $x$ to the characteristic coordinate $x + \gamma t$. We omit the details, but it is easily checked that the solution is,

$$J(x,t) = J(x+\gamma t, 0) + C\gamma^{-1}\left[\Phi(x) - \Phi(x+\gamma t)\right], \tag{17}$$

where

$$\Phi(x) \equiv \int_x^\infty \phi(x')\,dx'. \tag{18}$$

This function actually depends on the Voigt parameter $a$ (as does $\phi(x)$), but we again suppress this dependence in the notation. The first term of Eq. (17) gives the contribution from the initial values of $J$ at $t = 0$, while the second term gives the contribution from the constant source term.

For the purposes here, we are particularly interested in the case of zero initial field $J(x,0) = 0$. Then Eq. (17) can be written

$$J(x,t) = C\gamma^{-1}\left[\Phi(x) - \Phi(x+\gamma t)\right]. \tag{19}$$

This equation has a simple asymptotic solution for $x \gg 1$, where the profile function can be approximated by $\phi(x) \sim a/\pi x^2$, so that

$$J(x,t) \sim \frac{aCt}{\pi x(x+\gamma t)}, \qquad x \gg 1. \tag{20}$$

It will be shown later that this provides a good description of the asymptotic region $x \gg 1$ even when diffusion is present.

### 3.4. *Expanding Medium; Scaling Solution*

It is easy to write an approximate time-dependent transfer equation applicable to transfer in the Lorentz wings, in the spirit of Harrington (1973) and Basko (1979). We make the replacements $\phi(x) \sim a/\pi x^2$ in the diffusion term and use the delta-function injection approximation in the source term. Equation (11) then becomes

$$\frac{\partial J(x,t)}{\partial t} = \frac{a}{2\pi}\frac{\partial}{\partial x}\left(\frac{1}{x^2}\frac{\partial J}{\partial x}\right) + \gamma\frac{\partial J}{\partial x} + C\delta(x). \tag{21}$$

The time-independent (quasi-static) solution to this equation is very simple:

$$J(x,\infty) = C\gamma^{-1}\begin{cases} 1, & \text{if } x \le 0; \\ \exp(-2\pi\gamma x^3/3a), & \text{if } x \le 0, \end{cases} \tag{22}$$

(Chugai 1980). Note that the solution to the blue ($x \le 0$) is constant, but the solution to the red ($x \ge 0$) is cut off very sharply at a characteristic frequency of order $x_c = a^{1/3}\gamma^{-1/3}$.

We are unable to solve the full time-dependent equation (21) analytically; however, it is easy to derive some important scaling properties of the solution. Let us introduce the new scaled frequency and time variables $\xi$ and $\tau$ defined by

$$x = a^{1/3}\gamma^{-1/3}\xi, \qquad t = a^{1/3}\gamma^{-4/3}\tau. \tag{23}$$

It is straightforward to write Eq. (21) in terms of the new variables $\xi$ and $\tau$. With $F(\xi,\tau) \equiv C^{-1}\gamma J(x,t)$ this equation becomes,

$$\frac{\partial F(\xi,\tau)}{\partial \tau} = \frac{1}{2\pi}\frac{\partial}{\partial \xi}\left(\frac{1}{\xi^2}\frac{\partial F}{\partial \xi}\right) + \frac{\partial F}{\partial \xi} + \delta(\xi). \tag{24}$$

This equation no longer contains the parameters $a$, $\gamma$ and $C$, so its solution can be written as a function of the scaled variables $\xi$ and $\tau$ alone. These scaling properties may be emphasized by writing

$$J(x,t;a,\gamma,C) = C\gamma^{-1}F(\xi,\tau). \tag{25}$$

This form will be important for our discussion of relaxation time in the next section.

Note that Eqs. (21) and (24) only describe frequencies well outside the Doppler core. This implies that they are reasonable approximations only when the characteristic frequency scale $x_c = (\gamma/a)^{1/3}$ satisfies $x_c \gg 1$. Therefore the condition of validity for these equations is $\gamma \ll a$.

## 4. NUMERICAL RESULTS

The Fokker-Planck, diffusion-type equation represented by Eq. (11) was solved numerically using a straightforward generalization of the the Crank-Nicholson discretization scheme (see, e.g., Fletcher 1988; Press et al., 1992, p. 840), modified by the presence of the advection term $\gamma \partial J / \partial x$. Being semi-implicit in time, the Crank-Nicholson method has good stability properties. Most effort was concentrated on the pure constant source problem, $C = 1$, with zero inital field, $J(x, 0) = 0$, because of its importance to the relaxation time problem. The ranges of interest for the parameters $\gamma$ and $a$ for the Lyman alpha transition during recombination are roughly $-10 \leq \log \gamma \leq -6$ and $-3.5 \leq \log a \leq -2.5$. We actually studied a wider range of parameters that included the above ranges, namely, $-11 \leq \log \gamma \leq -1$ and $-6 \leq \log a \leq -1$, supplemented by the special cases $\gamma = 0$ (static case) and $a = 0$ (pure Doppler case).

### 4.1. *Static Medium*

As a test of our method, we applied it first to the static case ($\gamma = 0$), for which there are analytic solutions. Fig. 1 shows the time-dependent solution of the Fokker-Planck equation (11) for the special case of pure Doppler ($a = 0$), or $R_I$, redistribution. The exact Field solution (13) is also plotted for comparison (dashed lines). The Fokker-Planck equation is not expected to be a good description in this case, since the radiation field varies so rapidly with $x$. Indeed the agreement with the Field solution away from line center is very poor. However, the agreement near the line center is surprisingly good, since both solutions are very flat and of nearly the same level. It is comforting that, even in this most inauspicious case, the line center region (which determines the relaxation time) is not too badly represented by our improved Fokker-Planck equation.

Another static ($\gamma = 0$) test of the Fokker-Planck equation was a case for which $a = 10^{-3}$, plotted in Fig. 2. At early times, $t < 10^4$ (not shown), the solutions look very much like those of Fig. 1. At later times the solution begins to broaden considerably in the core region and to develop "wings," due to emission in the Lorentz portion of the Voigt profile. Eventually, for $t \gg 10^6$, the solutions near line center become very well represented by the approximate Basko solution (dashed lines) of Eq. (15). However, because of its delta-function injection approximation, the Basko solution does not match the solution in the region $|x| \gg x_c$, which is dominated by emission into the Lorentz wings.

### 4.2. *Expanding Medium*

In an expanding medium ($\gamma > 0$), the effect of the redshifting of the photons, described by the $\gamma(\partial J / \partial x)$ term in Eq. (11), eventually becomes important. The photons that scatter significantly outside the doppler core, where the time betweeen scatterings is large, are most affected by this redshifting. For those photons scattered blueward of line center, the effect of the expansion is to bring them back into the core of the line, where the scattering process can begin anew. Red photons, however, are redshifted further away from the line, and they are eventually lost. For asymptotically large times this leads to a time-independent solution (for a constant source) in which the injection of new photons near line center is balanced by the loss due to the redshift (see Rybicki and Hummer 1992).

Most of the preceeding effects can be seen in the simple (but otherwise unrealistic) case of coherent scattering. In Fig. 3 we plot the coherent solution (19) as $\log \gamma J$ vs. $x$ for $a = 10^{-3}$ and for various values of $\gamma t$. The line marked "$\infty$" is the quasi-static solution ($\gamma t \to \infty$), given by $\gamma J(x, \infty) = \Phi(x)$. Because of the scaling of $J$ and $t$, Fig. 3 applies to all values of $\gamma$. It is seen that for early times, $\gamma t \ll 1$, the solution is approximately equal to $\gamma t \phi(x)$ and increases linearly with time. As the solution approaches the quasi-static form at line center, $\gamma t \gtrsim 1$, the expansion term begins to dominate and the solution skews toward the red. Eventually the solution saturates to $\gamma J \sim 1$ in a broad region that extends from near line center to a limit in the red which continues to move at a roughly uniform rate, much like a wavefront.

The solution of the scaled Fokker-Planck equation (24) was found numerically using the Crank-Nicholson method. The results are plotted in Fig. 4. Due to the suppression of the Doppler core in the scaling equation, the early behavior is well described by the static Basko solution. At scaled times $\tau \sim 1$, the solution begins to skew toward the red, and at larger values of $\tau$, the solution approaches the time-independent Chugai solution (marked "$\infty$"). As in the case of the coherent solution, the late-time solution behaves like a front moving uniformly towards the red, but now this front is much less sharp. Also note the absence of emission in the Lorentz wings, because of the delta-function injection approximation implicit in equation (24).

We next found numerical solutions to the full $R_{II}$ redistribution problem, described by the Fokker-Planck equation (11). Solutions were obtained for a wide range of parameters, but detailed solutions will only be presented for a few representative examples, chosen for their relevance to the cosmological recombination problem. In Fig. 5 is plotted the numerical solution for a case where $a = 10^{-3}$ and $\gamma = 10^{-4}$. For early times, before the effects of expansion become important, the solution looks very much like those in Figs. 1 and 2, clearly showing the Doppler core. Later the expansion begins to skew the solution toward the red and eventually radiation is lost from the line and moves redward in the form of a front. Unlike the solution for the scaled equation in Fig. 4, Fig. 5 contains features due to emission in the Lorentz wings, and the asymptotic solution is still very much influenced by the Doppler core; in particular, the Chugai solution does not describe the asymptotic solution very well. The poor description in terms of the scaled Fokker-Planck equation (24) is explained by the only marginal satisfaction here of the condition $\gamma \ll a$.

In Fig. 6 is plotted a case where again $a = 10^{-3}$, but now $\gamma = 10^{-10}$, about the most extreme case of relevance to the cosmological problem. The frequency spreading is now very much larger than before, extending out to hundreds of Doppler widths, and the time scales are very much greater. The asymptotic solution is now well represented by the Chugai solution (22). In fact, the whole general behavior here is well represented by the scaled solution of Fig. 4, except for the very small emission in the wings.

In the very blue region $x \gg x_c$, the whole time dependent approach to the asymptotic solution is well described by the coherent solution (19), and in particular by the asymptotic form (19). This is because diffusion is here completely dominated by the redshifting.

## 5. THE RELAXATION TIME

### 5.1 *General Theory*

As explained in the Introduction, we are particularly interested in finding out when the quasi-static solution to the transfer problem is a good approximation to the full time-dependent equations. The quasi-static solution is defined as a time-independent solution with the values of the various slowly-varying parameters in the equations, such as $C(t)$, $\gamma(t)$, etc., frozen at their local values at some fiducial time. We can investigate this by determining the time required for an initially zero radiation field to become approximately equal to its asymptotic form in the constant parameter case, where $C(t) \equiv C$, $\gamma(t) \equiv \gamma$, etc. This defines a type of relaxation time, which we denote by $t_r$. If this relaxation time is much shorter than the characteristic times for the changes in the parameters, i.e., $C(t)/C'(t)$, $\gamma(t)/\gamma'(t)$, etc., then the quasi-static approach is valid.

Considerable care must be exercised in defining the criteria for the relaxation time. For example, from the above numerical solutions in Figs. 3–6 it can be seen that for any time, however large, there is always a sufficiently large negative frequency at which the solution is far from quasi-static, because there has not been sufficient time for radiation to redshift from the line center region. However, we note that the essential region of the quasi-static solution for recombination calculations is that near line center, which determines the trapping of the resonance radiation. We therefore define our relaxation time $t_r$ in terms of the time taken to reach the quasi-static solution at $x = 0$. To be more precise, we choose $t_r$ to be such that the radiation field at $x = 0$ reaches some predetermined fraction $f$ of the asymptotic value at $x = 0$. This definition can be formulated concisely through use of the *residual intensity* $r(t)$, defined as

$$r(t) = 1 - J(0,t)/J(0,\infty). \tag{26}$$

Then the relaxation time $t_r$ is determined by the condition

$$r(t_r) = f. \tag{27}$$

The choice of the fraction $f$ is somewhat arbitrary. As we shall see, the approach to the asymptotic solution is approximately exponential, so one natural choice might be $f = 1/e = 0.368$. However, we prefer the somewhat more conservative choice $f = 0.1$, so that $t_r$ is the time at which the time-dependent solution is within 10% of its quasi-static, asymptotic value, or, put another way, the quasi-static solution gives 10% accuracy. The exponential behavior then implies that the quasi-static solution gives 1% accuracy at two relaxation times, 0.1% at three relaxation times, etc.

As a simple example, let us evaluate the relaxation time for the case of coherent scattering, using the solution (19). Since in this case $\gamma C^{-1} J(0,\infty) = \Phi(0) = 1/2$, the residual intensity is simply $r(t) = 2\Phi(\gamma t)$, and the scaled relaxation time $t_r$ is determined by the relation $f = 2\Phi(\gamma t_r)$. This depends on the Voigt parameter $a$, but for $a \ll 1$ the result is essentially the same as that for $a = 0$, the pure Doppler case, so that the condition reduces to the simple one $f = \mathrm{erfc}(\gamma t_r)$. This leads to the result

$$t_r = K_c \gamma^{-1}, \qquad \text{(coherent scattering)} \tag{28}$$

where $K_c$ is a constant that depends on the choice for $f$. For example, our choice $f = 0.1$ implies $K_c = 1.16$.

Now let us derive the relaxation time for the case of $R_{II}$ redistribution. If we assume that the line transfer can be well described by Eq. (21), then a simple derivation of the relaxation time follows from the scaling relation, Eq. (25). Using this relation, we may write the quasi-static solution as $J(x, \infty) = C\gamma^{-1}F(\xi, \infty)$, so that the ratio of the time-dependent solution at to the quasi-static solution at $x = 0$ takes the form $J(0, t)/J(0, \infty) = F(0, \tau)/F(0, \infty)$. Therefore the residual intensity is a function of the single variable $\tau$, and the relaxation time in terms of $\tau$ is determined by the condition $r(\tau_r) = f$. In Fig. 7 the residual intensity is plotted using the numerical solution of Eq. (24); for $\tau \gtrsim 1$, it is quite accurately represented by the exponential law $r(\tau) = 0.920 \exp(-0.663\tau)$. For any choice of $f$, the relaxation time in terms of $\tau$ can be immediately read from this plot, say $\tau_r = K$. Then from Eq. (23) it follows that the scaled relaxation time for $R_{II}$ redistribution is

$$t_r = K a^{1/3} \gamma^{-4/3}. \tag{29}$$

For our choice $f = 0.1$ we find $K = 3.35$.

The formula (29) depends on the correctness of Eq. (21), which was derived on the basis of several approximations: that the transfer process is dominated by diffusion in the Lorentz wings, and that all new photons are injected exactly at $x = 0$. Thus the region of validity of the formula needs to be checked by direct numerical computations.

In Fig. 8 we plot typical results the relaxation times found from numerical solution of the full Fokker-Planck equation (11), with our preferred choice $f = 0.1$. This plot is based on 126 independent solutions covering the range $-6 < \log a < -1$ and $-11 < \log \gamma < -1$. There is a region in the upper part of the diagram for which the relaxation times are well represented by the asymptotic formula (29) with constant $K = 3.35$, in agreement with that found for the scaled equation. Note that the lower envelope of the scaling region corresponds to the condition $\gamma \ll a$ discussed at the end of §3.4.

The relaxation times $t_r$ in Eqs. (28) and (29) are dimensionless times defined according to Eq. (6). The corresponding physical relaxation time $t_{pr}$ for coherent scattering may be written

$$Ht_{pr} = K_c \beta_T. \qquad \text{(coherent scattering)} \tag{30}$$

Here we have used the definition (9) for $\gamma$ and have defined $\beta_T = v_T/c$, the relativistic parameter for the thermal motion of the atom, for which typically $\beta_T \ll 1$. In fact, during the recombination epoch, $\beta_T \approx 2 \times 10^{-5}$. Since $H^{-1}$ is the local Hubble time, which is a typical expansion time for the medium, Eq. (30) shows that, if coherent scattering were an accurate approximation, the relaxation time will always be very much less than the expansion time.

Similarly, the physical relaxation time for $R_{II}$ may be written

$$Ht_{pr} = K\beta_T \left(\frac{a}{\gamma}\right)^{1/3}. \tag{31}$$

This differs from the result (30) for coherent scattering in having a different proportionality constant and, more importantly, in having the extra factor $(a/\gamma)^{1/3}$. It is this latter factor that raises the possibility that the relaxation time might be comparable with characteristic times for expansion.

### 5.2 *Application to the Recombination Epoch*

We now wish to compare the relaxation time found above with the typical recombination timescales during the cosmological recombination epoch. We used solutions for recombination given by Peebles (1968) as well as our own independent solutions to give the values of various physical parameters as a function of time (or redshift).

Let us introduce a number of time scales in addition to the recombination time defined above. First, the expansion time $t_{exp}$ is defined by

$$t_{exp} = -\left(\frac{d\ln n_b}{dt}\right)^{-1}, \tag{32}$$

where $n_b$ is the baryon density. This is simply related to the Hubble time $t_H = H^{-1}$ by $t_{exp} = t_H/3$. More relevant to the problem at hand is the time scale associated with the rapid recombination of hydrogen, since this determines the time scale of the function $C(t)$. We thus define a recombination time $t_{rec}$ by

$$t_{rec} = -\left(\frac{d\ln n_e}{dt}\right)^{-1}, \tag{33}$$

where $n_e$ is the electron density. This time is typically 10% of the Hubble time in the middle of the recombination phase, roughly where $z \approx 1100$. Another time of interest is the scattering time $t_s \equiv 1/kc$ (cf., Eq. [6]), which represents the free time for a photon between scatterings near line center.

The cosmological model treated here is defined by current values of the density parameter $\Omega$, the density parameter due to baryons $\Omega_b$, and the scaled Hubble constant $h = H_0/(100 \text{ km s}^{-1} \text{ Mpc}^{-1})$. The cosmological constant is assumed to be zero; in any case this should not affect our results. In Fig. 9 we compare various time scales for the recombination problem for the set of cosmological parameters $\Omega = 1$, $\Omega = 0.06$, and $h = 0.5$. For this case it is seen that the recombination time is always greater than the relaxation time, but the ratio between them is as small as $\sim 20$ for $z \approx 900$.

The case we studied that gives the smallest factor between these two times, while still being fairly "reasonable," has $\Omega = 0.2$, $\Omega_b = 0.06$, and $h = 0.5$. The results for this case are shown in Fig. 10. Here the ratio between the recombination time and relaxation time is $\sim 5$ for a wide range of redshifts, $1000 \lesssim z \lesssim 1200$.

Because the relaxation time is uniformly smaller than the expansion time for all of our cosmological models, we conclude that a quasi-static treatment of the line transfer in Lyman alpha is an adequate approximation. The smallest factor between found between these times was 5, which, given the exponential approach found in §5.1, suggests that the accuracy of the quasi-static solution will be at least as good as one part in $10^5$.

## 6. OTHER PHYSICAL EFFECTS

Not included in the above theory is the slight change in frequency during scattering due to recoil of the atom. This effect was discussed qualitatively by Adams (1971), and a simple, analytic correction to the scattering term was derived by Basko (1981). Making Basko's approximations for our scattering term, we obtain the following alteration of Eq. (11),

$$\frac{\partial J}{\partial t} = \frac{1}{2}\frac{\partial}{\partial x}\left(\phi\frac{\partial J}{\partial x} + 2\eta\phi J\right) + \gamma\frac{\partial J}{\partial x} + C(t)\phi. \tag{34}$$

Here $\eta$ is the mean frequency shift in $x$ per scattering due to recoil, given by $\eta = h\nu_0/(Mcv_T)$. For the parameters of interest in the hydrogen recombination problem, we have $\eta \approx 5 \times 10^{-4}$. By approximating $\partial J/\partial x \approx J/x$, we see that the additional term in Eq. (34) only becomes important for frequencies $|x| \gtrsim (2\eta)^{-1} \approx 1000$, which do not play a significant role in determining the relaxation time. At large times, when the radiation field is spread to large red frequencies, the redshifting terms in the equation dominate over the small recoil effect. Thus, our neglect of this effect is justified. This has been confirmed numerically by solving the equations including the additional term for a number of cases covering the parameter range of interest.

A potentially more important effect is that the Lyman alpha scattering process can be interrupted, not by collisions (which are negligible for the cosmological problem), but by the ambient thermal radiation field (CMB) acting on atoms. During Lyman alpha scattering hydrogen atoms alternate between the $1S$ and $2P$ levels, but exist in the $2P$ level only very briefly ($\sim 10^{-9}$ s), almost always returning to the $1S$ level. However, while in the $2P$ level, the atom can be radiatively excited to some higher level by the CMB field. The subsequent cascade may bring the atom into the metastable $2S$ level, from which it most likely would be excited again by the CMB, but some fraction could undergo two-photon decay to $1S$. The two-photon loss mechanism would tend to reduce the relaxation time, since it shortens the scattering process. On the other hand, if the atom is returned to the $2P$ level, so that Lyman alpha scattering continues, it will start by being emitted in the Voigt profile, not as a continuation of $R_{II}$ redistribution. This might change the solution by putting more radiation further out in the wings of the line, delaying relaxation. Such CMB radiative transitions are expected to be quite unlikely for any given scattering event, but might nonetheless affect the transfer process because of the large numbers of scatterings experienced by a typical Lyman alpha photon. The evaluation of this effect will be left to future work.

## 7. CONCLUSIONS

We have investigated the time-dependent radiative transfer in a resonance line with $R_{II}$ redistribution, which is appropriate to the hydrogen Lyman alpha line during cosmological recombination. A new, improved Fokker-Planck approximation was derived, and several new analytic results were presented. One principal result is the derivation of the analytic formula, Eq. (29), giving the scaled relaxation time for the line radiation field to approach its quasi-static form for $R_{II}$ redistribution. This formula is written in terms of physical variables in Eq. (31). Extensive numerical solutions of the Fokker-Planck transfer equation were performed in order to normalize and check this relaxation formula.

In order to test if quasi-static models for the line transfer are adequate during cosmological recombination, we compared the relaxation time with characteristic time scales for recombination, as determined from detailed models of the recombination epoch. For all reasonable cosmological parameters it was found that the relaxation time is uniformly smaller than the recombination time, but in some extreme cases by a factor of no more than 5. We conclude that that reasonably accurate results can be obtained using quasi-static solutions as the basis for cosmological recombination calculations. This is quite fortunate, since time-dependent calculations involving both radiation and matter would be much more complicated.

## APPENDIX
## ANGLE DEPENDENT FOKKER-PLANCK OPERATOR FOR $R_{II}$

In this appendix we derive an improved Fokker-Planck approximation for the $R_{II}$ scattering operator. The derivation is actually somewhat more general than required for this paper, in that we treat the case of an angle-dependent radiation field. The isotropic case follows as a special case.

The three-dimensional form of the scattering operator for line scattering with $R_{II}$ redistribution in static media can be written

$$\frac{1}{4\pi}\int d\Omega' \int dx'\, R_{II}(x,\mathbf{n};x',\mathbf{n}')I(\mathbf{r},x',\mathbf{n}'), \qquad (A1)$$

where $R_{II}(x,\mathbf{n};x',\mathbf{n}')$ is the angle dependent redistribution function (see, e.g., Hummer 1962).

In the Fokker-Planck method the integral operator defining the scattering term is replaced by an second-order differential operator. This method is appropriate when the frequency width of the redistribution function is small compared to the scale of frequency variation of the intensity. This is precisely the situation for $R_{II}$ redistribution for transfer in the Lorentz wings of the line, and we expect that a Fokker-Planck approach will work well there. In fact, Fokker-Planck type equations for $R_{II}$ redistribution were derived and used by Unno (1955) and Harrington (1973). However, the equations derived by these authors were restricted to the asymptotic wing regime, and become formally singular in the core of the line. Furthermore, these equations assumed the angle-averaged form of the $R_{II}$ redistribution function.

It is the purpose here to provide a derivation of a simple new Fokker-Planck equation for $R_{II}$ redistribution, one based on exact evaluations of the first three moments of the complete angle-dependent redistribution function, after Frisch & Bardos (1981). However, it is shown that all of these exact moments cannot be used simultaneously, because they are inconsistent with photon conservation upon scattering. By introducing a slight alteration of the second moment, we obtain a simple Fokker-Planck operator that does conserve photons. In its angle-averaged form, this new operator is identical to that used by Harrington in the wings of the line, but has the advantage of being nonsingular in the core.

In order to reduce the scattering term to a differential operator we expand the $x'$ dependence of $I(\mathbf{r}, x', \mathbf{n}')$ in a power series about $x$:

$$I(\mathbf{r}, x', \mathbf{n}') = I(\mathbf{r}, x, \mathbf{n}') + (x' - x)\frac{\partial}{\partial x}I(\mathbf{r}, x, \mathbf{n}') + \frac{1}{2}(x' - x)^2\frac{\partial^2}{\partial x^2}I(\mathbf{r}, x, \mathbf{n}') + \cdots. \quad (A2)$$

Then to second order,

$$\frac{1}{4\pi}\int d\Omega' \int dx' \, R_{II}(x, \mathbf{n}; x', \mathbf{n}')I(\mathbf{r}, x', \mathbf{n}') = \frac{1}{4\pi}\int d\Omega' \Big[A_0(x, \mathbf{n}, \mathbf{n}')I(\mathbf{r}, x, \mathbf{n}') + A_1(x, \mathbf{n}, \mathbf{n}')\frac{\partial}{\partial x}I(\mathbf{r}, x, \mathbf{n}') + \frac{1}{2}A_2(x, \mathbf{n}, \mathbf{n}')\frac{\partial^2}{\partial x^2}I(\mathbf{r}, x, \mathbf{n}')\Big], \quad (A3)$$

where

$$A_k(x, \mathbf{n}, \mathbf{n}') = \int dx' \, (x' - x)^k R_{II}(x, \mathbf{n}; x', \mathbf{n}'). \quad (A4)$$

To evaluate the $A_k$ we make use of the double Fourier transform representation of $R_{II}$,

$$\tilde{R}_{II}(\tau, \mathbf{n}; \tau', \mathbf{n}') \equiv \int_{-\infty}^{\infty} dx \int_{-\infty}^{\infty} dx' \, e^{i\tau x + i\tau' x'} R_{II}(x, \mathbf{n}; x', \mathbf{n}'). \quad (A5)$$

It has been shown (Rybicki, unpublished; Heinzel 1981) that this can be expressed in the closed form

$$\tilde{R}_{II}(\tau, \mathbf{n}; \tau', \mathbf{n}') = g(\mathbf{n}\cdot\mathbf{n}')e^{-\frac{1}{4}\tau^2} e^{-\frac{1}{4}\tau'^2} e^{-a|\tau+\tau'|} e^{-\frac{1}{2}\tau\tau'\mathbf{n}\cdot\mathbf{n}'}. \quad (A6)$$

Here $g(\mathbf{n}\cdot\mathbf{n}')$ is the phase function for the scattering, which can be taken to be isotropic (case A) or dipole (case B),

$$g_A(\mathbf{n}\cdot\mathbf{n}') = 1; \qquad g_B(\mathbf{n}\cdot\mathbf{n}') = \frac{3}{4}\left[1 + (\mathbf{n}\cdot\mathbf{n}')^2\right]. \quad (A7)$$

The result (6) can be used to evaluate the Fourier transform of $A_k$:

$$\tilde{A}_k(\tau, \mathbf{n}, \mathbf{n}') \equiv \int_{-\infty}^{\infty} dx \, e^{i\tau x} A_k(x, \mathbf{n}, \mathbf{n}'). \quad (A8)$$

First of all we write

$$\tilde{R}_{II}(\tau - \tau', \mathbf{n}; \tau', \mathbf{n}') \equiv \int_{-\infty}^{\infty} dx \, e^{i\tau x} \int_{-\infty}^{\infty} dx' \, e^{i\tau'(x - x')} R_{II}(x, \mathbf{n}; x', \mathbf{n}'), \quad (A9)$$

so that

$$\tilde{A}_k(\tau, \mathbf{n}, \mathbf{n}') = \left(\frac{\partial}{i\partial\tau'}\right)^k \tilde{R}_{II}(\tau - \tau', \mathbf{n}; \tau', \mathbf{n}')\Big|_{\tau'=0}. \quad (A10)$$

Since we need $A_k$ for only the few values $k = 0, 1, 2$, the easiest way to calculate them is by expanding $\tilde{R}_{II}(\tau - \tau', \mathbf{n}; \tau', \mathbf{n}')$ to second order in $\tau'$,

$$\begin{aligned}\tilde{R}_{II}(\tau - \tau', \mathbf{n}; \tau', \mathbf{n}') &= g(\mathbf{n}\cdot\mathbf{n}')e^{-\frac{1}{4}\tau^2 - a|\tau|} e^{\frac{1}{2}\tau\tau'(1-\mathbf{n}\cdot\mathbf{n}')} e^{-\frac{1}{2}\tau'^2(1-\mathbf{n}\cdot\mathbf{n}')}, \\ &= g(\mathbf{n}\cdot\mathbf{n}')e^{-\frac{1}{4}\tau^2 - a|\tau|}\Big\{1 + \frac{1}{2}\tau(1 - \mathbf{n}\cdot\mathbf{n}')\tau' \\ &\quad - \frac{1}{2}\left[(1 - \mathbf{n}\cdot\mathbf{n}') - \frac{1}{4}\tau^2(1 - \mathbf{n}\cdot\mathbf{n}')^2\right]\tau'^2 + \cdots\Big\}. \end{aligned} \quad (A11)$$

Thus we read off the results

$$\begin{aligned}
\widetilde{A}_0(\tau,\mathbf{n},\mathbf{n}') &= g(\mathbf{n}\cdot\mathbf{n}')e^{-\frac{1}{4}\tau^2-a|\tau|},\\
\widetilde{A}_1(\tau,\mathbf{n},\mathbf{n}') &= -\frac{1}{2}i\tau g(\mathbf{n}\cdot\mathbf{n}')(1-\mathbf{n}\cdot\mathbf{n}')e^{-\frac{1}{4}\tau^2-a|\tau|},\\
\widetilde{A}_2(\tau,\mathbf{n},\mathbf{n}') &= g(\mathbf{n}\cdot\mathbf{n}')\left[(1-\mathbf{n}\cdot\mathbf{n}')-\frac{1}{4}\tau^2(1-\mathbf{n}\cdot\mathbf{n}')^2\right]e^{-\frac{1}{4}\tau^2-a|\tau|}.
\end{aligned} \tag{A12}$$

Recognizing that $\exp(-\frac{1}{2}\tau^2 - a|\tau|)$ is the Fourier transform of the Voigt profile $\phi(x)$ and that multiplication by $(-i\tau)$ is equivalent to $\partial/\partial x$, we have the exact results,

$$\begin{aligned}
A_0(x,\mathbf{n},\mathbf{n}') &= g(\mathbf{n}\cdot\mathbf{n}')\phi(x),\\
A_1(x,\mathbf{n},\mathbf{n}') &= \frac{1}{2}g(\mathbf{n}\cdot\mathbf{n}')(1-\mathbf{n}\cdot\mathbf{n}')\phi'(x),\\
A_2(x,\mathbf{n},\mathbf{n}') &= g(\mathbf{n}\cdot\mathbf{n}')(1-\mathbf{n}\cdot\mathbf{n}')\phi(x)+\frac{1}{4}g(\mathbf{n}\cdot\mathbf{n}')(1-\mathbf{n}\cdot\mathbf{n}')^2\phi''(x),
\end{aligned} \tag{A13}$$

where the primes here denote differentiation with respect to $x$.

Although we now have found exact expressions for $A_k$, $k = 0, 1, 2$, it is very important *not* to use the exact expressions for $A_1$ and $A_2$ simultaneously. The reason is that the differential operator that represents the right hand side of the transfer equation should conserve particles in the conservative scattering case; this requires it to be a divergence, which in turn requires that

$$A_1 = \frac{1}{2}A_2'. \tag{A14}$$

We see that the exact values of $A_1$ and $A_2$ do not obey this requirement. In other words, the expansion (A2) and (A3) leads to coefficients that do not ensure exact photon conservation when truncated to a finite series.

Experience with radiative transfer problems involving large numbers of scatterings has shown that even small errors in photon conservation can lead to very large errors in the solution. On the other hand, if exact photon conservation is satisfied, small errors in the coefficients lead only to small errors in the solution. Therefore, we shall make Eq. (A14) exact by making appropriate alterations in either $A_1$ or $A_2$, or both. A careful examination of the exact coefficients (A13) shows that one particularly simple choice stands out, namely to adopt the exact expression for $A_1$, but to approximate $A_2$ by its first term,

$$A_2(x,\mathbf{n},\mathbf{n}') = g(\mathbf{n}\cdot\mathbf{n}')(1-\mathbf{n}\cdot\mathbf{n}')\phi(x). \tag{A15}$$

We note that this is an excellent approximation in the asymptotic region $|x| \gg 1$, where $\phi''/\phi \sim 6/x^2$. Near the core, where $x \lesssim 1$, we simply accept the "error" in $A_2$ as being unavoidable in any Fokker-Planck treatment that preserves photon conservation. It would perhaps be of some interest to try other alterations of $A_1$ and $A_2$ to see whether they offer any special advantages. However, the simplicity of the present choice recommends it highly, and we shall consider no other here.

For coefficients satisfying the relation (A14), the scattering term (A3) can be expressed in the form

$$\begin{aligned}
&\frac{1}{4\pi}\int d\Omega' \int dx'\, R_{II}(x,\mathbf{n};x',\mathbf{n}') I(\mathbf{r},x',\mathbf{n}') \\
&= \frac{1}{4\pi}\int d\Omega' \left[A_0(x,\mathbf{n},\mathbf{n}') I(\mathbf{r},x,\mathbf{n}') + \frac{1}{2}\frac{\partial}{\partial x}\left(A_2(x,\mathbf{n},\mathbf{n}')\frac{\partial}{\partial x} I(\mathbf{r},x,\mathbf{n}')\right)\right]. \quad (A16)
\end{aligned}$$

Using our adopted values for $A_k$ the scattering term (A16) can now be displayed explicitly. For example, in the isotropic scattering case,

$$\begin{aligned}
&\frac{1}{4\pi}\int d\Omega' \int dx'\, R_{IIA}(x,\mathbf{n};x',\mathbf{n}') I(\mathbf{r},x',\mathbf{n}') \\
&= \phi(x) J(\mathbf{r},x) + \frac{1}{2}\frac{\partial}{\partial x}\left\{\phi(x)\frac{\partial}{\partial x}\left[J(\mathbf{r},x) - n_i H_i(\mathbf{r},x)\right]\right\}
\end{aligned} \quad (A17)$$

and for the dipole case,

$$\begin{aligned}
\frac{1}{4\pi}\int d\Omega' \int dx'\, R_{IIB}(x,\mathbf{n};x',\mathbf{n}') I(\mathbf{r},x',\mathbf{n}') &= \phi(x)\frac{3}{4}\left[J(\mathbf{r},x) + n_i n_j K_{ij}(\mathbf{r},x)\right] \\
+ \frac{3}{8}\frac{\partial}{\partial x}\left\{\phi(x)\frac{\partial}{\partial x}\left[J(\mathbf{r},x) + n_i n_j K_{ij}(\mathbf{r},x) - n_i H_i(\mathbf{r},x) - n_i n_j n_k L_{ijk}(\mathbf{r},x)\right]\right\}&, \quad (A18)
\end{aligned}$$

where the moments of the radiation field are defined by

$$\begin{aligned}
J(\mathbf{r},x) &= \frac{1}{4\pi}\int I(\mathbf{r},x,\mathbf{n})\, d\Omega, \\
H_i(\mathbf{r},x) &= \frac{1}{4\pi}\int I(\mathbf{r},x,\mathbf{n}) n_i\, d\Omega, \\
K_{ij}(\mathbf{r},x) &= \frac{1}{4\pi}\int I(\mathbf{r},x,\mathbf{n}) n_i n_j\, d\Omega, \\
L_{ijk}(\mathbf{r},x) &= \frac{1}{4\pi}\int I(\mathbf{r},x,\mathbf{n}) n_i n_j n_k\, d\Omega.
\end{aligned} \quad (A19)$$

For notational convenience, an obvious tensor notation (and summation convention) has been introduced.

In many physical situations, e.g., the cosmological case, the radiation field is isotropic, and the above formulas simplify greatly, since the odd moments vanish, $H_i = 0$ and $L_{ijk} = 0$, and by symmetry the second moment is given by $K_{ij} = (1/3)\delta_{ij} J$. In this case the scattering term has the same form for either the isotropic or dipole phase function,

$$\int dx'\, R_{II}(x,x') J(\mathbf{r},x') = \phi(x) J(\mathbf{r},x) + \frac{1}{2}\frac{\partial}{\partial x}\left(\phi(x)\frac{\partial J(\mathbf{r},x)}{\partial x}\right). \quad (A20)$$

It is this form that is used in the body of the paper.

## FIGURE CAPTIONS

Figure 1. Static solutions ($\gamma = 0$) for pure Doppler redistribution ($a = 0$), found from numerical solution of the Fokker-Planck equation (11). The dashed curves are the exact analytic results of Field (1959).

Figure 2. Static solutions ($\gamma = 0$) for $R_{II}$ redistribution with $a = 10^{-3}$. The solid curves are the numerical results for the Fokker-Planck equation (11) and the dashed curves are based on the analytic Basko solution (Eq. [15]).

Figure 3. Plot of Eq. (19) for coherent scattering with expansion for $a = 10^{-3}$. The use of scaled variables $\gamma J(x)$ and $\gamma t$ make this plot valid for all values of $\gamma$.

Figure 4. Numerical solution of the scaled Fokker-Planck equation (24). Here the scaled frequency and time variables are $\xi = a^{-1/3}\gamma^{1/3}x$ and $\tau = a^{-1/3}\gamma^{4/3}t$. This plot applies to any values of $a$ and $\gamma$ within the region of validity of the equation (see text).

Figure 5. Numerical solution of the Fokker-Planck equation (11) for $a = 10^{-3}$ and $\gamma = 10^{-4}$. The times are given by values of $t_5 \equiv t/10^5$. The curves below the $10^{-3}$ curve correspond to $t_5 = 10^{-4}$, $10^{-5}$, etc. The asymptotic, quasistatic solution is marked "$\infty$."

Figure 6. Same as Fig. 5, except $\gamma = 10^{-10}$ and times are given by values of $t_{12} \equiv t/10^{12}$. The curve marked "C" is the Chugai solution (22).

Figure 7. The residual intensity $r(\tau)$ vs. $\tau$ for the scaled Fokker-Planck equation (24). The straight dashed line (barely seen in the upper left) is the exponential fit to the lower part of the curve.

Figure 8. Log-log plot of the relaxation time $t_r$ vs. $\gamma$ for various values of $\log a$. These values are based on numerical solution of Eq. (11) with the choice $f = 0.1$.

Figure 9. Plot of various time scales (see text) during recombination epoch with cosmological parameters $\Omega = 1$, $\Omega_b = 0.06$, and $h = 0.5$.

Figure 10. Same as Fig. 9, except with $\Omega = 0.2$

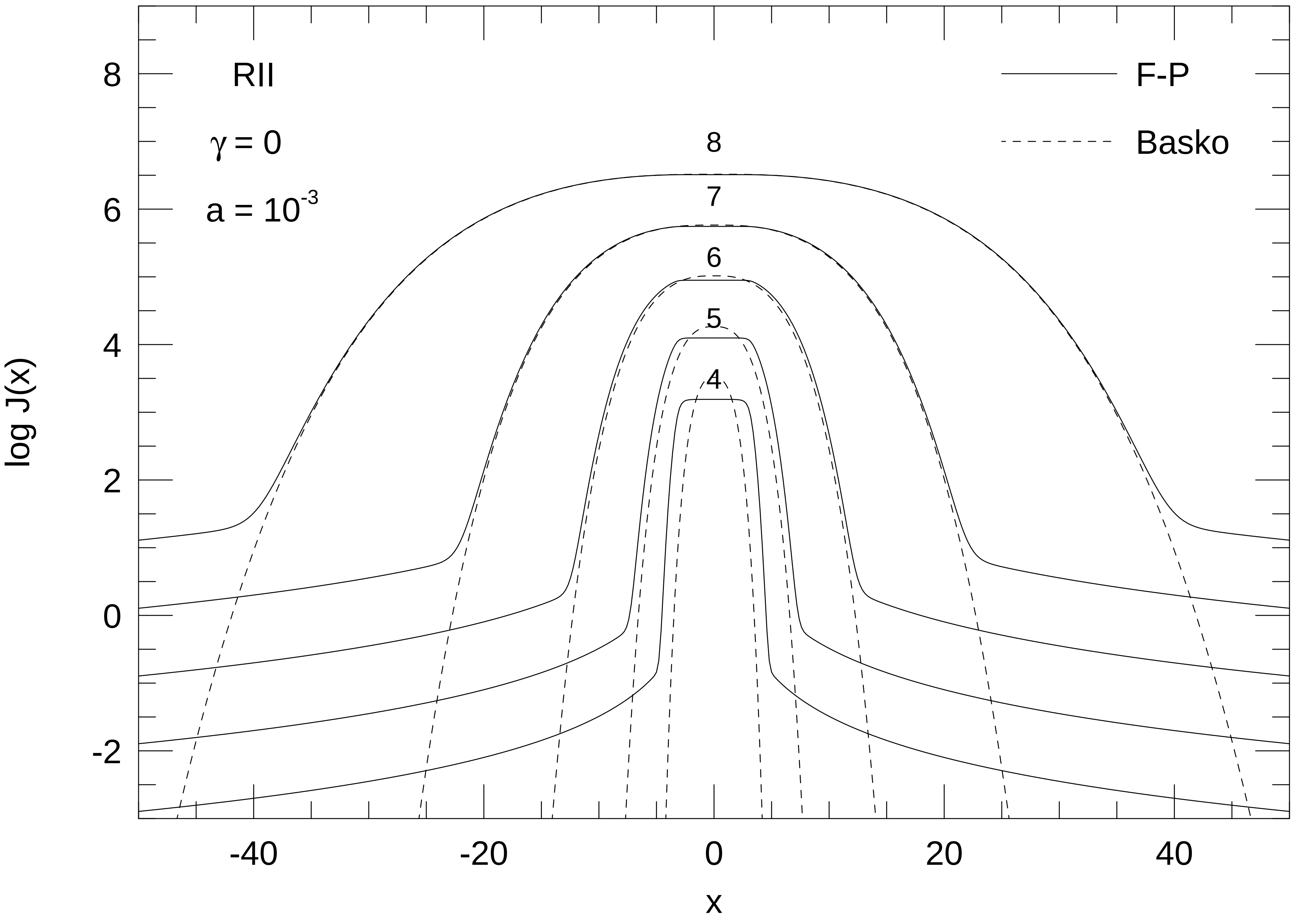

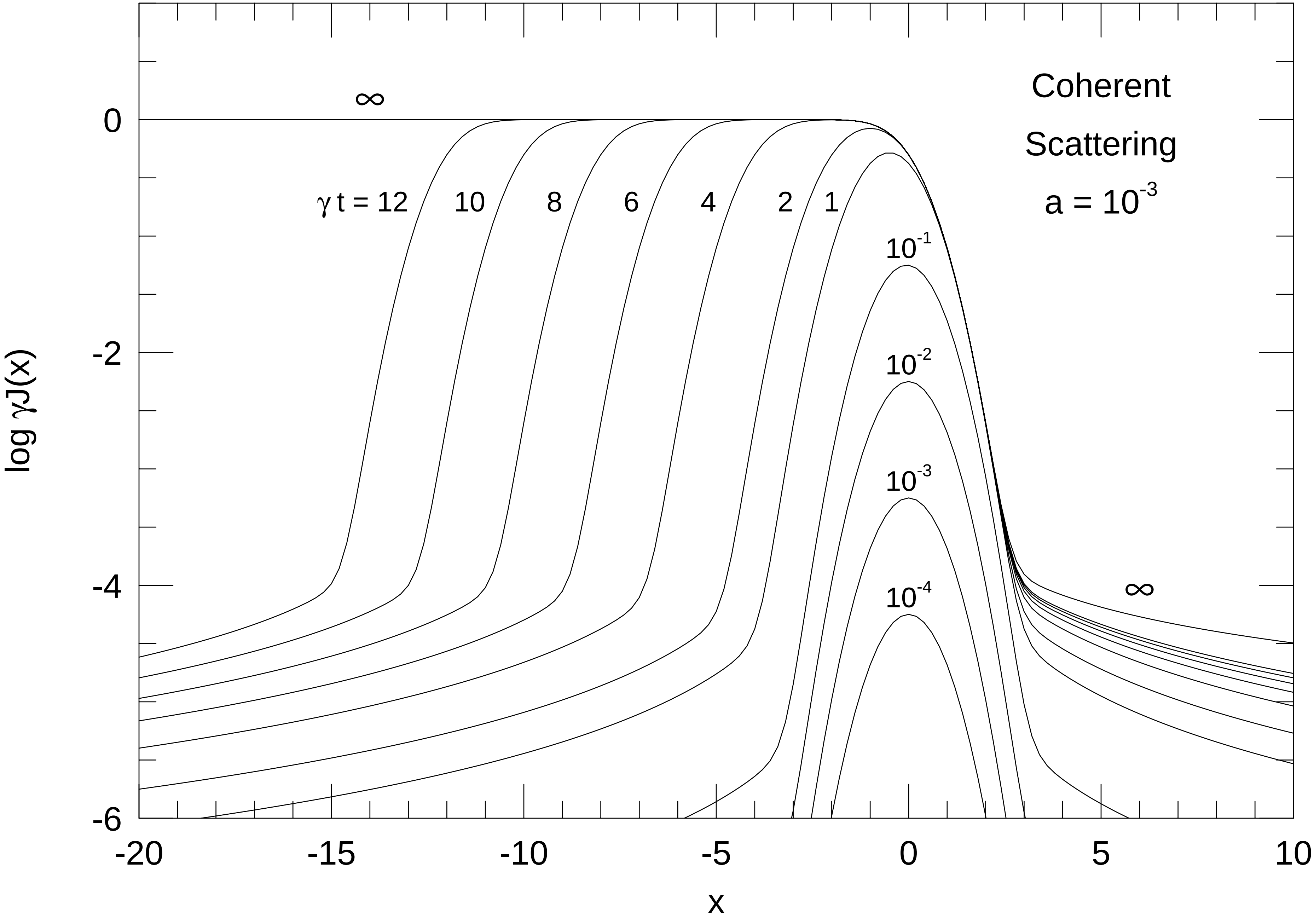

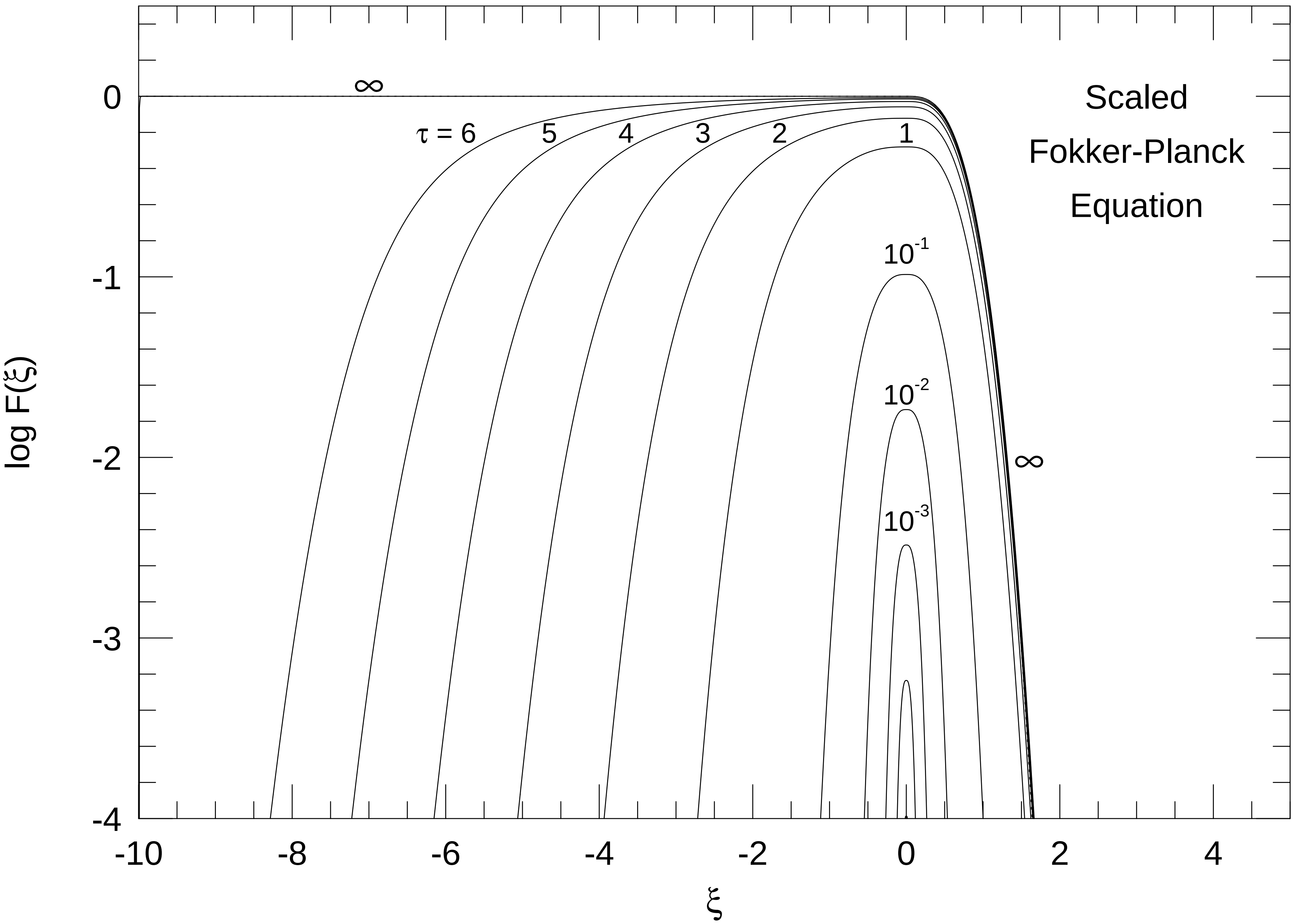

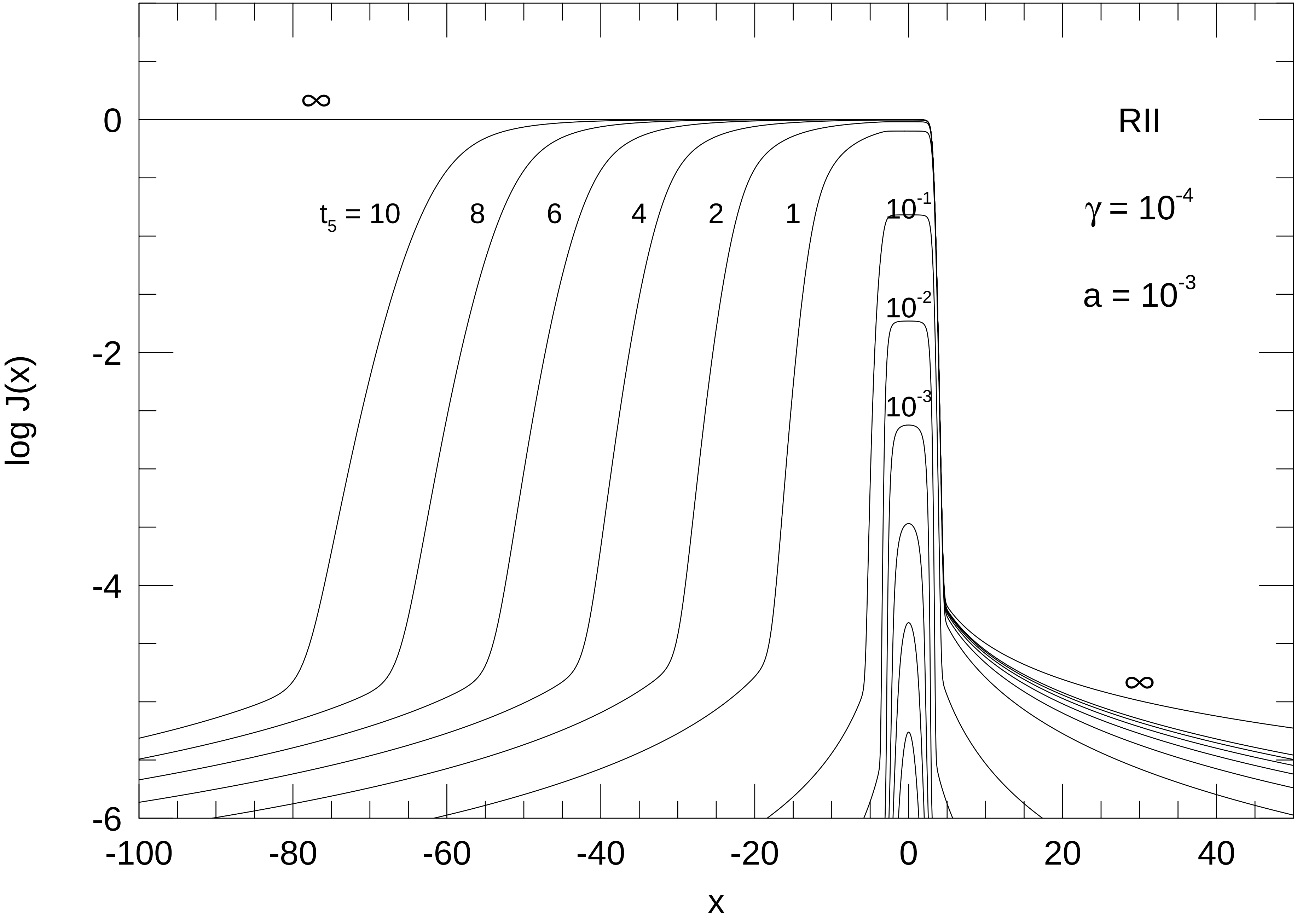

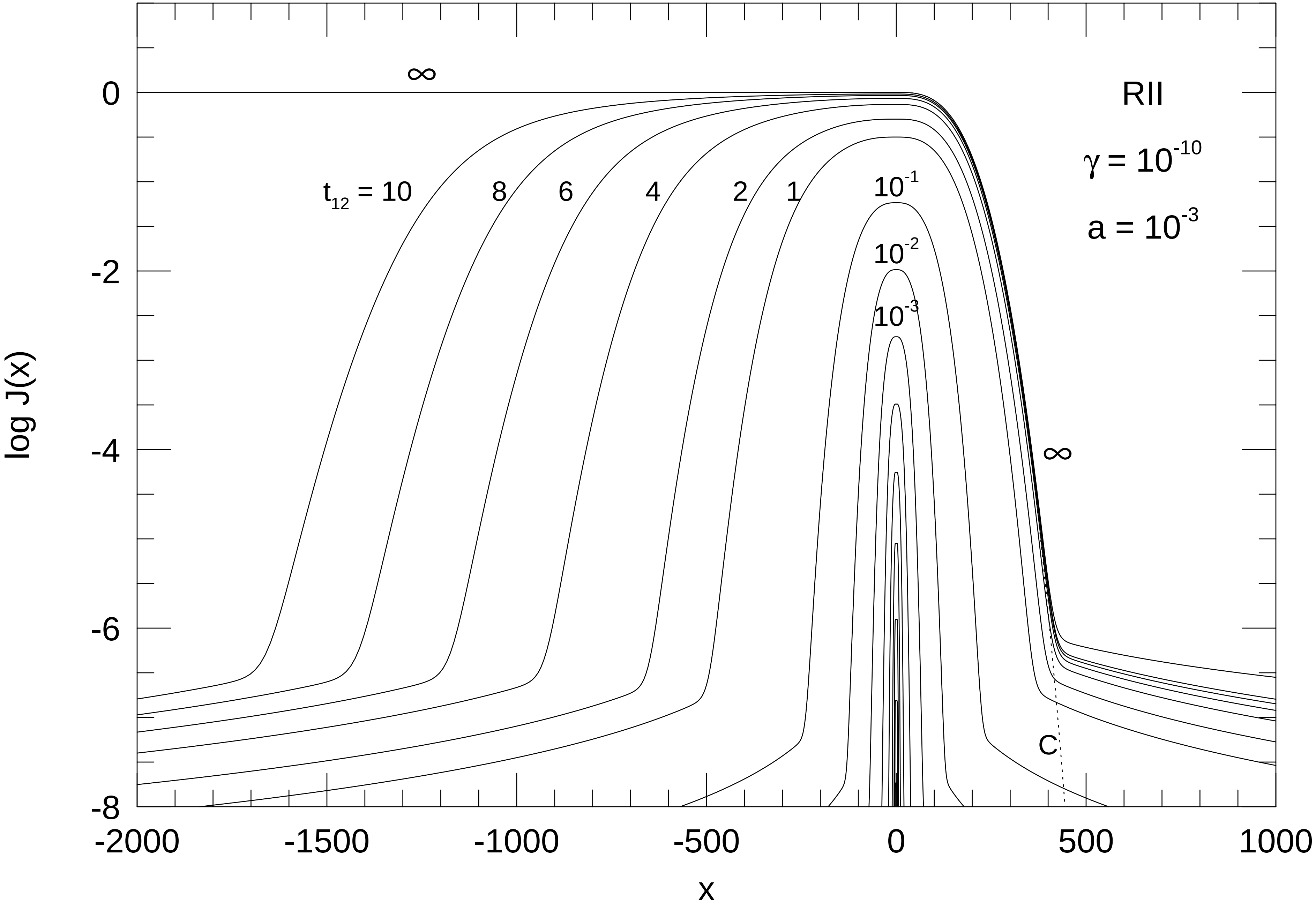

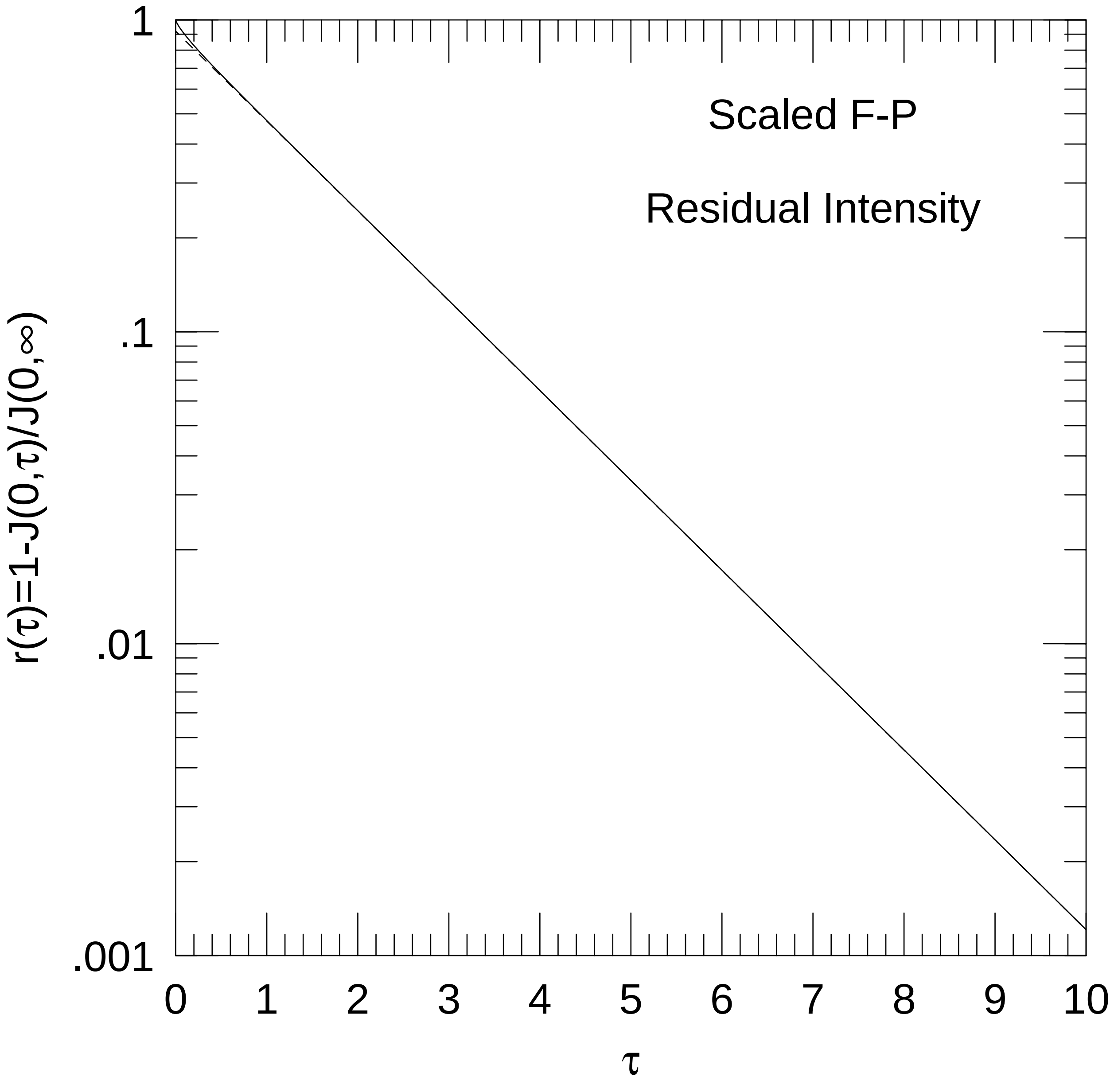

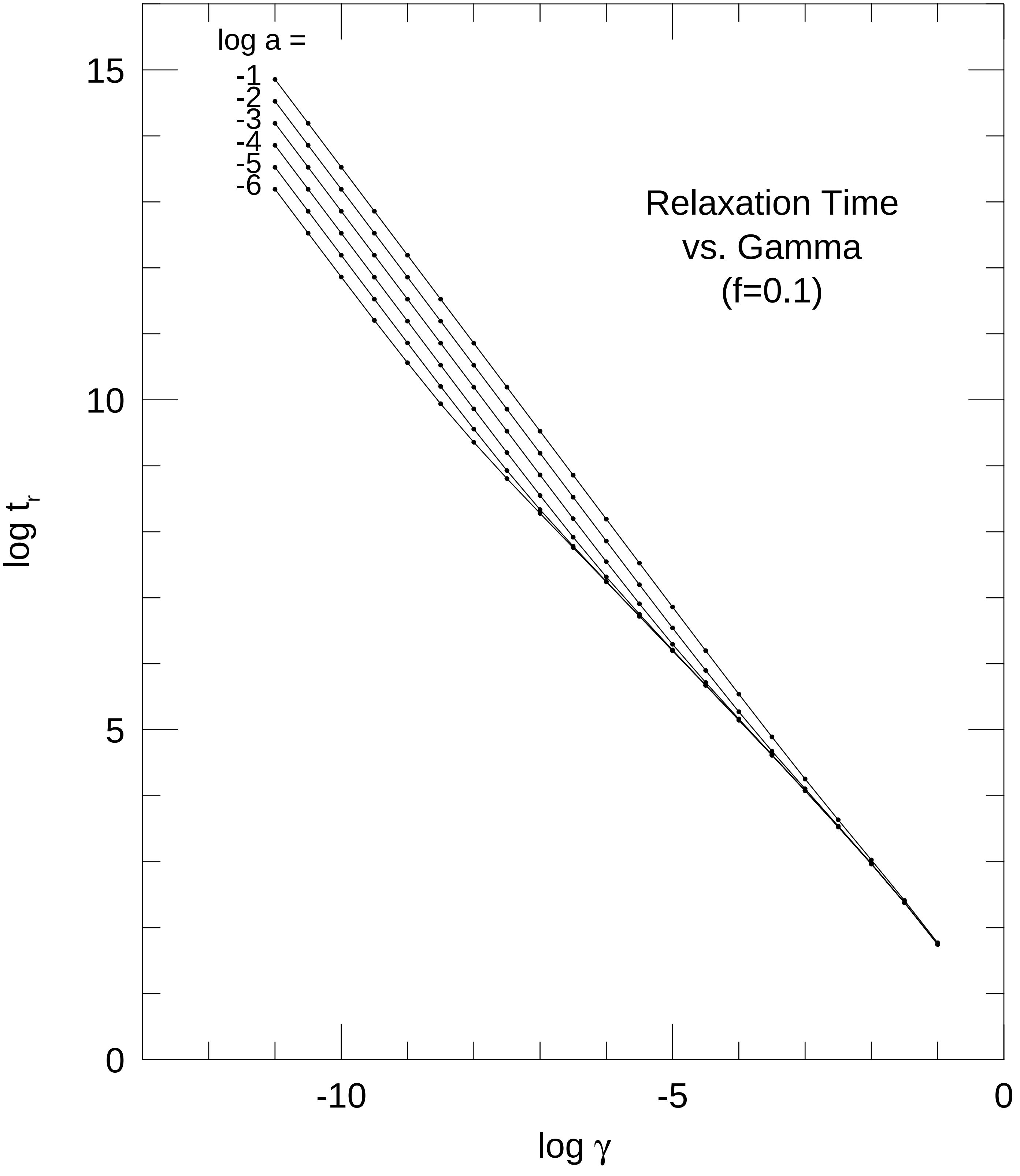

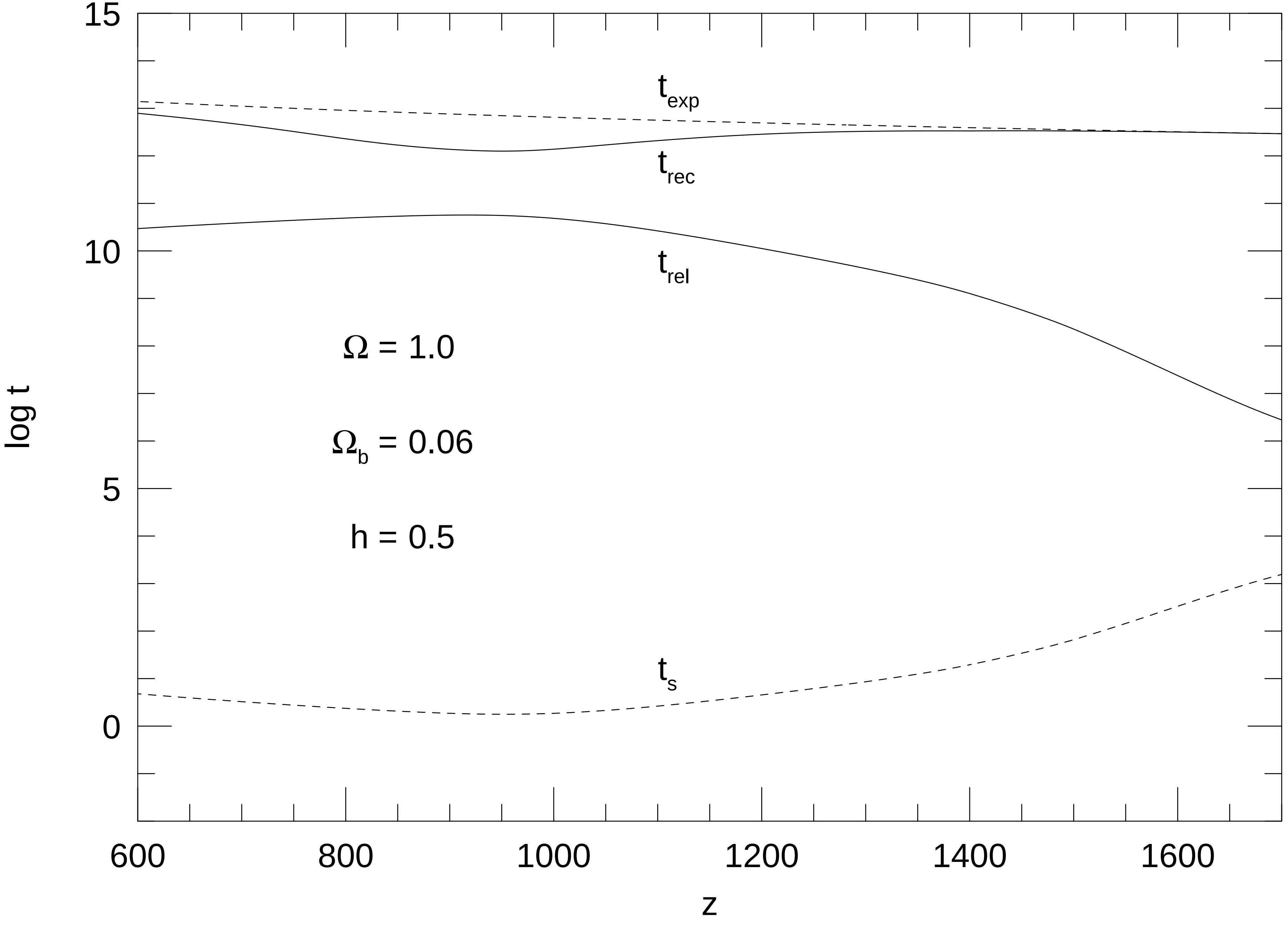

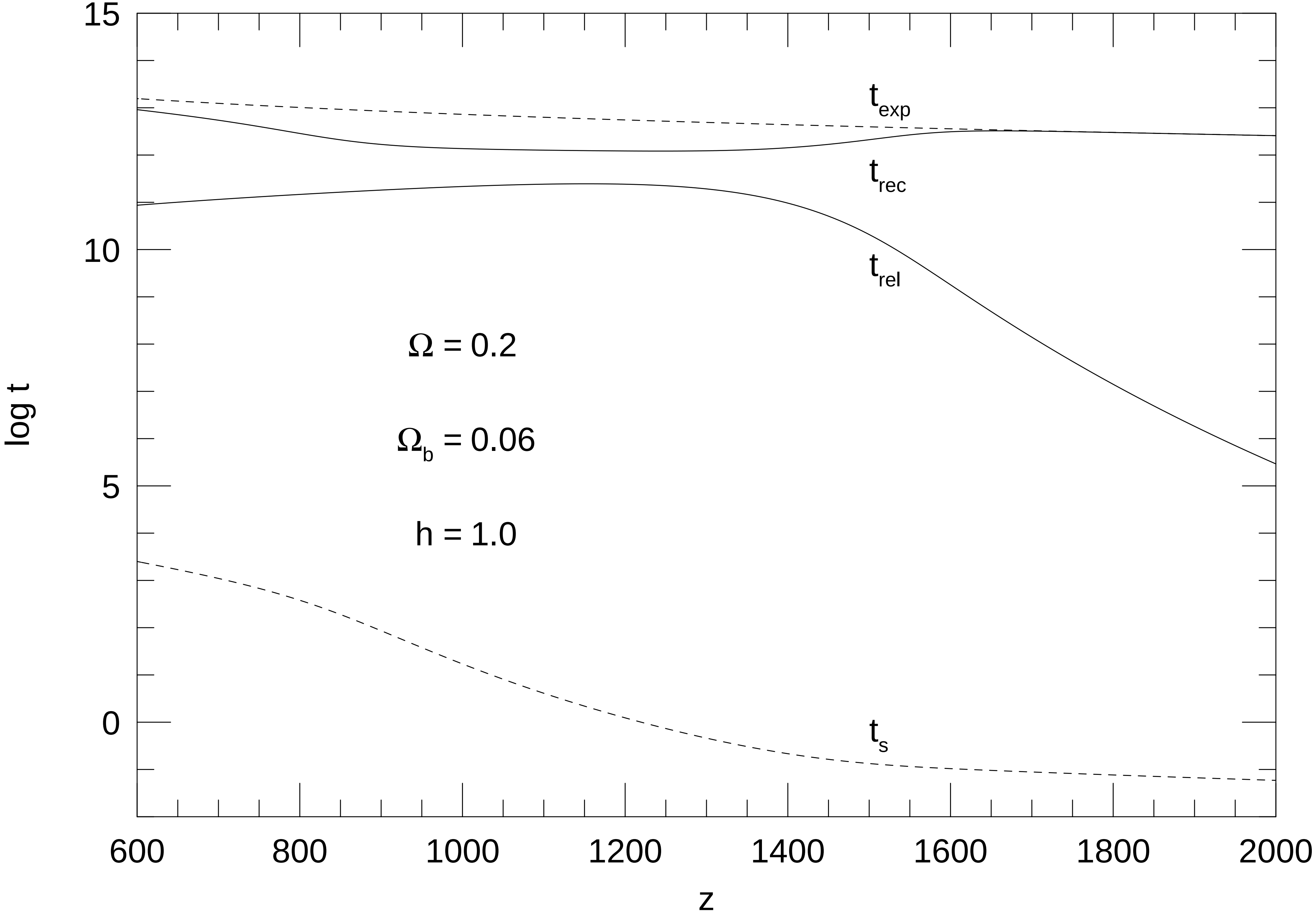